# Design and dynamics experiment of filter wheel mechanism of space coronagraph

Wei Guo[a,b], Jiangpei Dou[a,b], Mingming Xu*[a,b], Linyi Kong[a,b], Baolu Liu[a,b], Bo Chen[a,b], Shu Jiang[a,b]

[a] National Astronomical Observatories / Nanjing Institute of Astronomical Optics & Technology, Chinese Academy of Sciences, Nanjing 210042, China;

[b] CAS Key Laboratory of Astronomical Optics & Technology, Nanjing Institute of Astronomical Optics & Technology, Nanjing 210042,China;

**ABSTRACT**

In order to realize multi-spectral imaging of space coronagraph, a compact filter wheel mechanism is designed. The filters with different spectral transmittance can be cut into the optical path at different times by this mechanism. Because the image contrast of space coronagraph is very high, the high stability requirement for the optical unit of coronagraph is put forward. Small modulus worm gear and worm are taken by the mechanism, in order to realize compact structure, high stiffness, the unidirectional 360° rotation and reverse self-locking function. The precision, stiffness, mechanical properties and reliability of the mechanism are analyzed in the paper. The results show that the position accuracy of the filter wheel can meet the requirement of ≤±0.5mm. The first order modal of the mechanism is 313Hz. The results of vibration test indicate that stiffness, dynamic performance and reliability of the mechanism can be meet. Therefore, the design of filter wheel in this paper can ensure the multi-spectral imaging requirements under complex spatial conditions

## 1. INTRODUCTION

Is life on the Earth unique in the universe? Are there habitable worlds outside our solar system that could support life? These are the questions that mankind has been searching for and looking forward to answer for a long time [1]. Extrasolar planet detection is a hot research topic in international astronomy nowadays,which has become one of the hottest research directions in the field of astronomy[2].

Space coronagraph exoplanet detection technology can achieve high contrast, wide band, large detection area and other advantages. The scientific goal of coronagraph is to conduct high contrast imaging detection study for the exploration of extrasolar planets and break through the limitation of the imaging contrast detection ability of ground observation equipment and provide ultra-high contrast imaging.

Filter wheel mechanism is an important part of scientific imaging for exoplanet exploration, which is the main space motion unit of the device. It is necessary to design an optical splitter with high precision and high stability for space coronagraph in order to ensure the imaging quality of multi-spectral segments. The filter wheel can be cut in and out through the rotational motion of the transmission mechanism, which can realize the subdivision of the spectrum. Because the filter wheel is close to the scientific imaging camera, it was required to switch frequently during the lifetime of the orbit. The filter wheel mechanism is usually a single point of failure in the equipment, so the design of mechanism with simple structure, long life, high stiffness and high reliability is the premise to ensure the clear observation of exoplanets in orbit.

Corresponding author: Mingming Xu , xmm4544@sina.com

The James Webb Space Telescope (JWST) is equipped with a low-temperature filter wheel mechanism [3-4] and a near infrared channel cryogenic filter wheel [5], which are both center driving structure. Although they have been applied to mid-wave infrared spectrum and near-infrared spectrum successfully, their filter diameters are small and the structure size is big. Too much space is occupied on the optical path direction and the stiffness requirement is high. They are not suitable for compact layouts. Huili Jia, et al. develop a large size thin wall bearing filter wheel for GF-4, which can filter different spectral segments [6-8]. Because the mechanism ensures the self-locking of the components by the self-positioning torque of the motor, it is necessary to achieve load locking by amplifying the self-positioning torque of the motor with large deceleration ratio. The reducer structure size is relatively large on the radial direction. He Bao et al. design a new type of filter wheel mechanism based on the base of a pentagonal pyramid, which can increase the structural stiffness of the filter wheel [9]. High precision, high strength, high reliability and self - locking function can be achieved by the mechanism. Because the connection tension of the structure is generated due to the unbalanced installation between the base of the pentagonal pyramid and the filter wheel components, the high assembly precision of structure is required.

Because of the high contrast requirements for exoplanet imaging, the compact structure and high reliability and high stiffness and dynamic performance of filter wheel are required.

According to the requirements of space coronagraph, a compact and ultra-stable filter wheel mechanism based on small modulus worm gear and worm driving is designed in this paper, which is used to detect space astronomy. Function and performance and precision are analysed in the paper. The reliability of the mechanism was verified by modal analysis and sinusoidal and random vibration tests in the paper.

## 2. DESIGN OF SPACE FILTER WHEEL MECHANISM

The filter wheel mechanism is equipped with four optical filters, which are B1(910-970nm), B2(1100-1200nm), B3(1350-1500nm) and B4(1495-1781nm). Narrow-band optical filters in the near infrared band are installed for imaging band selection in scientific observations. When an exoplanet target is observed, the optical filter is rotated to a specific band position for measuring celestial objects accurately by motion mechanism. The diameter of the optical filter is 25.4mm and the optical aperture is 16mm. The switching time between adjacent spectrum segments must be less than 10 seconds. The filter wheel speed is initially set at 3RPM. The optical filter eccentricity error is less than ±0.5mm. The Overall weight is less than 1.8kg.

The motor static torque margin must be more than 1 in order that the working performance of the mechanism in orbit can be ensured. The first-order modal of the mechanism must be more than 100Hz in order that the complex mechanical environment performance during launch can be meet and low frequency resonance between the mechanism and the rocket during launch can be avoided. According to the space coronagraph observation model, the mechanism will work for ten years in orbit.

Combining with the design requirements of filter wheel, a compact design scheme of filter wheel mechanism is presented in this paper. The stepper motor is used to drive. The Small modulus worm gear worm are used for transmission mechanism. An encoder is used for position feedback and a Hall device is used for a backup feedback channel. Four optical filters are uniform distribution on the turnplate. The turntable of filter wheel is connected with the output end of the worm gear. The input of worm is connected with the output shaft of stepper motor. The transmission ratio of worm gear and worm is 48. When the turntable is rotated by worm drive, the position is monitored by the encoder. When the encoder sends a feedback signal, the closed position loop can be realized. The product design structure is shown in Figure 1.

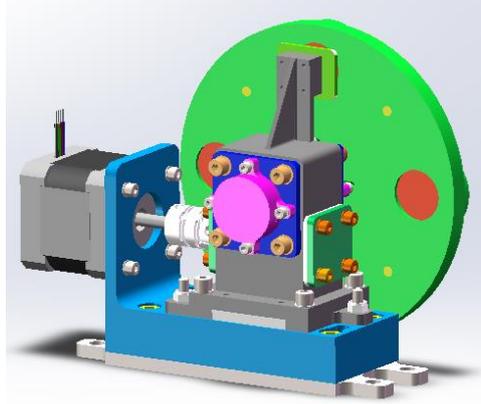

Figure.1 Structure design scheme of Filter wheel

## 3. THE PARAMETER DESIGN AND PRECISION ANALYSIS OF FILTER WHEEL

### 3.1 Motor driving parameters and mechanism self-locking performance analysis

J45BYG005A stepper motor is selected as the driving component for the filter wheel device. The pulling moment of stepper motor is more than 0.16N·m (starting frequency $f$=125Hz). The load inertia moment of the system $M_1$ is 10 N·m. The rolling friction torque of deep groove ball bearing $T_1$ is 24×10$^{-3}$N·mm. The friction torque of angular contact ball bearing $T_2$ is equal to 53.83×10$^{-3}$N·mm. Since there is no gravity in space, the drag moment is zero. The stable operation torque $M_2'$ is calculated as follow:

$$M_2' = 2T_1 + \frac{2T_2}{\eta i} = 0.3544 N \cdot mm, \qquad (1)$$

The transmission efficiency of worm gear and worm η is 0.4. The drive margin of motor Q is calculated as follow:

$$Q = \frac{\eta T}{R} - 1 = \frac{\eta T}{M_1 + M_2'} - 1 = 5.37 > 1, \qquad (2)$$

The requirements of driving margin can be meet.

The filter wheel must withstand the complex dynamics of rocket launch in the transportation and launch phase. Thus, mechanism must have capacity of certain locking capability in order to prevent the filter assembly from moving. At the same time, when the scientific observation of exoplanets is carried out by the scientific payload in orbit, the filter wheel needs to keep the stationary and locked state for imaging observation when it rotates to a certain band position. Therefore, the self-locking of the mechanism is very important for the normal operation.

The positioning torque of the motor T1 is more than 0.01N·m. The acceleration of the mechanism is less than or equal to 10g during the launch phase. Thus, the self-locking torque margin of the motor is:

$$\frac{\eta T_0 i}{(10T_0 - T_1)} - 1 = 1.13. \qquad (3)$$

The result shows that the self-positioning torque of the motor is more than 1. The locking function of the component can be meet in the launch stage.

The lead angle of small modulus worm γ is 2°52′. The worm is made of 9Cr18, which is quenching and tempering treatment. The worm gear is made of QSn6.5-1. Molybdenum disulfide solid lubrication is used between the worm gear and worm. The friction coefficient μ is equal to 0.08~0.1.The friction coefficient is set 0.08 in the paper. The friction Angle is

calculated as follow:

$$\rho_v = \arctan \mu = 4.57°, \quad (4)$$

Because the equivalent friction angle is more than the worm lead Angle, the reverse self-locking property of worm gear and worm drive is reliable.

4 filters are evenly arranged on the turntable of the mechanism. The switching angle of single optical filter $\theta_1$ is equal to 90°. The operating frequency of motor f is 400 Hz. The stepping angle is equal to $\theta=1.8°$. Then the switching time is calculated as follow:

$$t = \frac{\theta_1 i}{f_2 \theta} = 6s \leq 10s, \quad (5)$$

Because the switching time of optical filter of the adjacent spectral segment is less than 10 seconds, which can meet requirement.

### 3.2 Precision analysis of filter wheel mechanism

#### 3.3.1 Positioning precision analysis of optical axis of filter wheel mechanism

The distance between the optical axis of the filter and the shaft of the worm gear is 45mm. Then the optical axis positioning precision of the filter wheel mechanism is:

$$\varepsilon = \theta \cdot \frac{1}{i} \cdot R \cdot \frac{2\pi}{360} = 0.03\text{mm}. \quad (6)$$

The positioning precision of the optical axis of the mechanism is better than 0.5mm.

#### 3.3.2 Calculation of rotation precision of filter wheel mechanism

The machining errors and assembly errors are inevitable in the process of part manufacturing and assembly. Manufacturing errors of parts, assembly errors and transmission errors will directly affect the positioning precision of the filter. The main errors that affect the precision of the wheel mechanism include transmission error of worm gear and worm, step angle error of stepper motor, bearing error and measurement error of position sensor.

The step error of stepper motor $\theta$ is 1.8°±5%. Then the angular error of the rotation of the filter wheel $\delta_1$ is:

$$\delta_1 = \frac{\Delta\theta}{i} = \frac{0.09°}{48} = 0.001875°. \quad (7)$$

Machining precision of worm gear and worm is 6 level tolerance. The limit deviation of worm gear tooth pitch $f_{pt}$ is ±0.007mm. The limit deviation of worm gear tooth pitch $f_{pt}$ is calculated as 0.01mm. The transmission errors of worm gear and worm $\delta_2$ is calculated as follow:

$$\delta_2 = \arctan(\frac{0.01}{24}) = 0.024°. \quad (8)$$

The precision of bearing is 4 grade. The shafting error $\delta_3$ is 0.02°. The reference radius of worm gear r is 12mm. The estimated null return error between worm gear and worm is 0.02mm。Then the Angle error $\delta_4$ is calculated as follow：

$$\delta_4 = \arctan(\frac{0.02}{r}) = \arctan(\frac{0.02}{12}) = 0.097°. \quad (9)$$

Then the comprehensive error value is calculated as：

$$\delta = \sqrt{\delta_1^2 + \delta_2^2 + \delta_3^2} + \frac{\delta_4^2}{\sqrt{2}}$$
$$= \sqrt{0.001875^2 + 0.024^2 + 0.002^2} + \frac{0.097°}{\sqrt{2}} \qquad (10)$$
$$= 0.075°.$$

The precision of 18-bit encoder $\zeta$ is 25 "(0.0069°), When the encoder is used for position feedback, the rotation error is:
$$\delta + \zeta = 0.075° + 0.0069° = 0.0819°.$$
The distance between the optical axis of the filter and the shaft of the worm gear R is 45mm, The corresponding maximum linear error of the center is:
$$45 \times tg(0.0819°) = 0.064 \text{mm}. \qquad (11)$$

If the encoder fails in orbit, the backup Hall switch will be enabled for position monitoring feedback. The measurement precision of the Hall device used in the filter wheel is about 0.15mm. The distance between the Hall device and the rotating shaft is 45mm. The Angle error is 0.382°.

The rotation error of filter wheel mechanism and the measurement error of Hall device are considered comprehensively. The Error of mechanism is :
$$0.075° + 0.382° = 0.457°. \qquad (12)$$
The corresponding maximum linear error is:
$$45 \times tan(0.457°) = 0.359 \text{mm} < \pm 0.5 \text{mm}. \qquad (13)$$
Therefore, the rotation precision of the filter runner meets the eccentricity error requirement .

## 4. STIFFNESS ANALYSIS AND DYNAMIC PERFORMANCE ANALYSIS OF FILTER WHEEL

### 4.1 Modal analysis of mechanism

According to the weight and stiffness requirements of the filter wheel mechanism, the bracket, reducer box and turntable are made of TC4。 The worm is made of 9Cr18.The worm gear is made of QSn6.5-1.The optical filter is made of fused quartz. The optical filter size is φ25.4mm×4mm. The filters were mounted in titanium frames and fixed with epoxy glue. The pressing ring made of POM is used to fix the optical filter in the axial direction. In order that the center of mass of the turntable can be coincided with the worm gear axle and the additional torque caused by the rotation center of mass offset can be reduced ,Roundness and symmetry structure design of turntable should be ensured as far as possible. The assembly model of filter wheel mechanism is shown in Figure2.

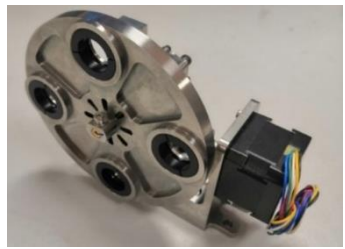

Figure.2 Assembly model of filter wheel mechanism

The low order modal of spatial mechanism is an important index to investigate dynamic stiffness. If the lower modal overlaps or are close to the natural modes of the rocket or satellite, the mechanism will resonate. Resonance is easy to reduce the stiffness of the mechanism and even leads to the failure of the mechanism [10]. The dynamic performance of the filter wheel mechanism is analyzed by different prediction methods, which is an effective method for dynamic analysis of spatial structure [11]. The finite element model of the mechanism was established, which is shown in Figure 3. It contains 188,917 units and 462,398 nodes. The connecting interface of each component includes screw connection, bearing fit and rubber bonding. The first four step modal shapes are shown in Figure 4 and the modal frequency results are shown in Table 1.

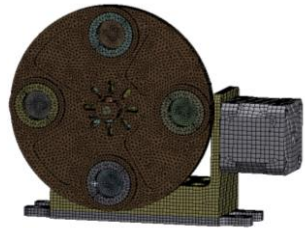

Figure. 3 Finite element model

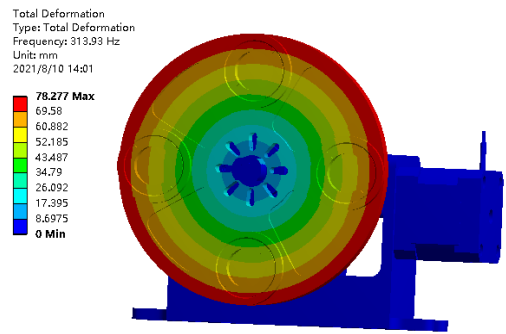

(a) The first-order modal shape

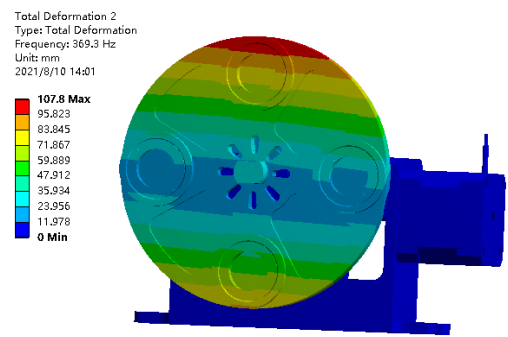

(b) The second-order modal shape

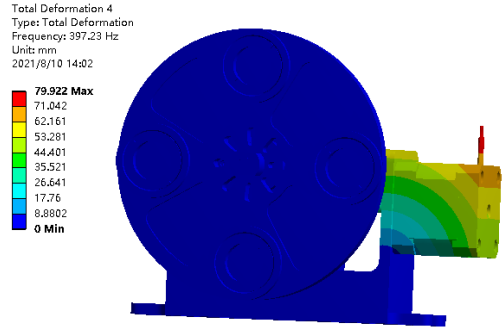

(c) The third-order modal shape

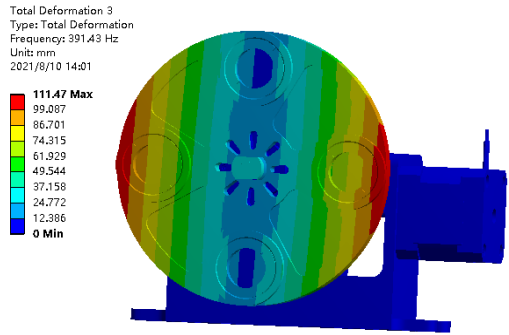

(d) The fourth-order modal shape

Figure. 4 Modal analysis of filter wheel mechanism

Table 1 The first to fourth orders modal of Filter wheel

| order | 1 | 2 | 3 | 4 |
|---|---|---|---|---|
| frequency/Hz | 313 | 369 | 391 | 397 |

The first-order modal of the mechanism is 313Hz, which is more than the resonant frequency of the rocket. The stiffness requirements of the mechanism can be meet.

**4.2 Mechanical properties analysis of emission environment of filter wheel mechanism**

The filter wheel mechanism must withstand the harsh dynamic environment in the process of transmitting. It is necessary to examine the dynamic characteristics of the filter wheel mechanism under complex conditions, such as sinusoidal vibration and random vibration. Sufficient mechanical simulation analysis of the mechanism is required to verify the anti-vibration and impact ability of the components. The rocket will vibrate due to the insufficient combustion of the booster during the launch. Although the dynamic load environment is aperiodic in the actual process of launch, periodic sinusoidal load is mostly used for the analysis of low-frequency vibration load below 100Hz. Vibration response analysis was carried out in X, Y and Z directions. DC-3200-36/SV-0606 vibrostand is used to perform sinusoidal and random vibration tests. The filter wheel is installed on the vibrostand, which is shown in Figure 5. Four acceleration sensors A1, A2, A3 and A4 are used to monitor response. Two acceleration sensors B1 and B2 are used as control points, which are pasted on the vibrostand.

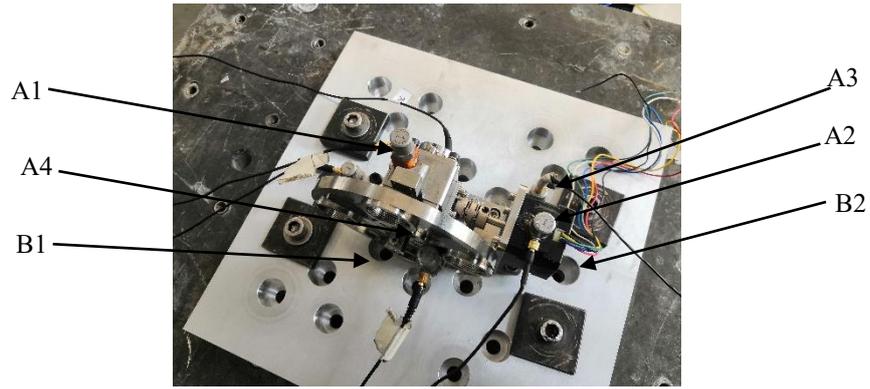

Figure.5 Vibration environment test

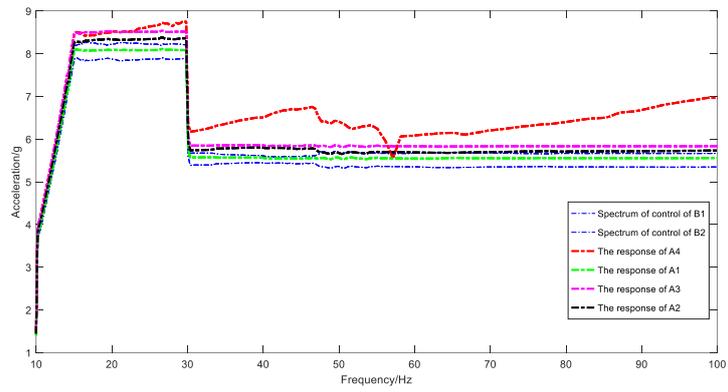

(a) Sinusoidal vibration test results in X direction of the filter wheel

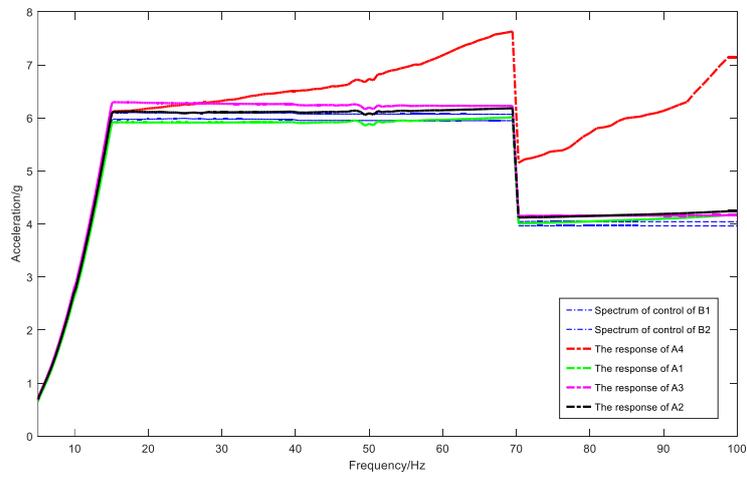

(b) Sinusoidal vibration test results in Y direction of the filter wheel

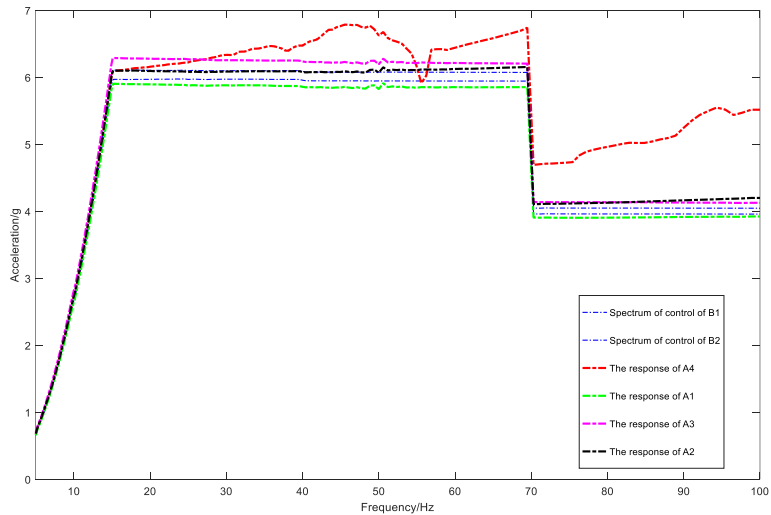

(c) Sinusoidal vibration test results in Z direction of the filter wheel

Figure.6 Sinusoidal vibration test results of filter wheel

The sinusoidal vibration test curves of the filter wheel in three directions are shown from Figure 6(a) to 6(c). The maximum response in the X direction is at the A4 monitor (filter turntable). When the frequency is 29.5Hz, the acceleration amplitude is 8.749g. The resonance magnification is about 1.1 times. The maximum response in Y direction is at A4 monitoring (filter turntable). When the frequency is 69.5Hz, the acceleration amplitude is 7.63g. The resonance magnification is about 1.27 times. The maximum response in Z direction is at A4 monitoring (filter turntable). When the frequency is 45.5Hz, the acceleration amplitude is 6.7873g. The resonance magnification is about 1.13 times.

According to the analysis of sinusoidal vibration curves in three directions, there is no obvious resonance amplification in the filter wheel mechanism components within 5-100Hz. The maximum resonance magnification of the three directions is about 1.27 times. The micro resonance is caused by the local modal of the mechanism. The sinusoidal vibration test of the filter wheel mechanism shows that the mechanism has sufficient stiffness in the range of low frequency sinusoidal vibration.

The random vibration test curves of the filter wheel in three directions are shown from Figure 7(a) to 7(c). The maximum acceleration of total root mean square of random vibration is 12.859g in X direction. The maximum acceleration of total root mean square of random vibration is 17.4162g in Y direction. The maximum acceleration of total root mean square of random vibration is 24.0479g in Z direction.

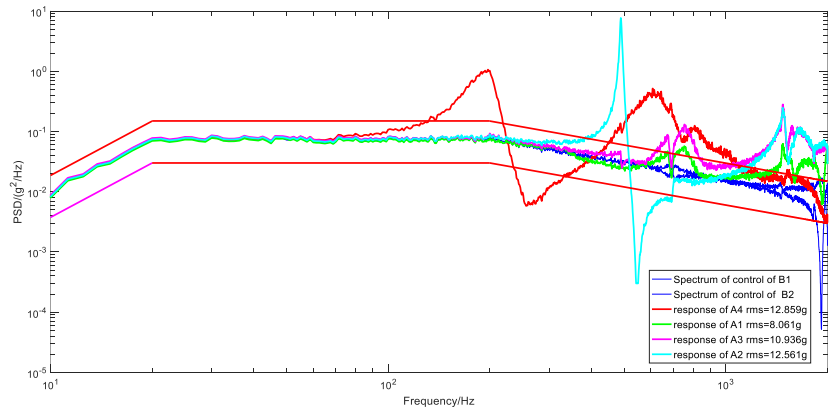

(a) Random vibration test results in X direction

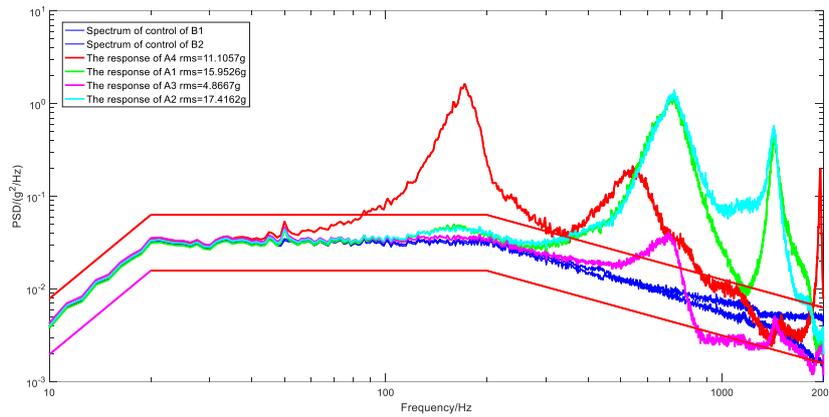

(b) Random vibration test results in Y direction

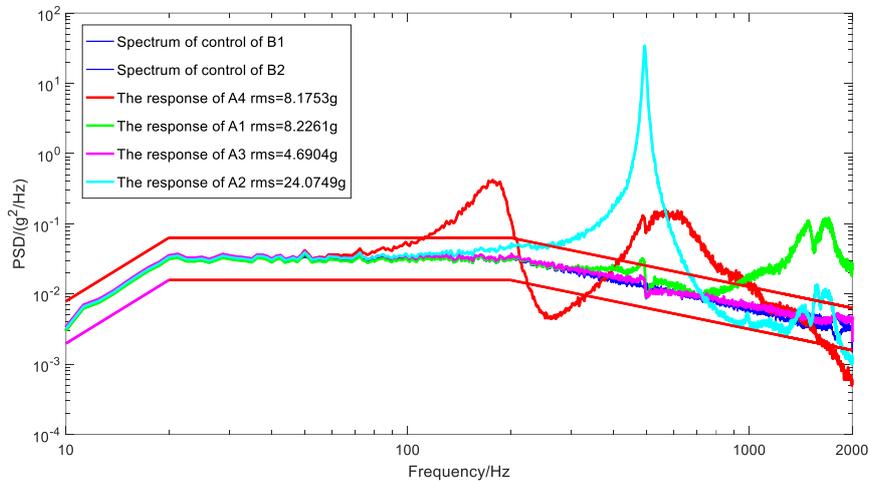

(c) Random vibration test results in Z direction

Figure. 7 Random vibration test results of filter wheel

The filter wheel works well after the sinusoidal and random vibration tests. The stiffness can withstand the mechanical environmental test of sinusoidal vibration and random vibration of qualification level. The mechanical requirements of rocket launch can be meet.

## 5. CONCLUSION

In order to realize ultra-high contrast multispectral imaging of exoplanets, a filter wheel device with high stiffness, compact structure and good self-locking function is designed in this paper. The driving moment margin and rotation precision of the mechanism were analyzed. The stiffness and vibration performance of the mechanism were analyzed and tested. The results show that:

1) The first-order natural frequency of the filter wheel mechanism is 313Hz, which meets the stiffness requirements of the mechanism.
2) The analysis result of driving margin of the system is 5.37, which meets the design torque requirements. The self-locking torque margin of the motor is 1.13. Because the lead angle of worm is $\gamma=2°52' < \rho_v$, reverse self-locking requirement can be meet.
3) When the encoder feedback is used, the positioning precision of the filter wheel mechanism is 0.078mm. When Hall switch feedback is used, the positioning precision of the filter wheel mechanism is 0.359mm. Two kinds of precisions are both better than 0.5mm. The overall precision can meet the requirement of design.
4) The dynamic characteristics of the filter wheel mechanism are obtained through the sinusoidal and random vibration tests. The results show that there is no obvious amplification of vibration magnitude, which can withstand the environmental test and meet the requirements of launch stage.

In conclusion, the space filter wheel mechanism developed in this paper can meet the precision requirements and the performance can meet the mechanical environment requirements. The multi-spectrum switching imaging under complex working conditions can be realized. The filter wheel mechanism scheme provided in this paper provides a reference for the design of the future high resolution multispectral phase camera for exoplanet exploration.

## ACKNOWLEDGMENTS

This work was supported by the National Natural Science Foundation of China (Grant No.12103073) and the project of China Space Station Engineering Survey Telescope Exoplanet Imaging Coronagraph.